\documentclass[preprint,12pt]{elsarticle}

\usepackage{graphics}

\usepackage{amssymb}

\journal{Physical Letters B}

\begin{document}

\begin{frontmatter}

\title{Higgs mass from cosmological and astrophysical measurements}

\author{L. A. Popa}

\address{Institutul de \c{S}tiin\c{t}e Spa\c{t}iale Bucure\c{s}ti-M\u{a}gurele, Ro-077125 Rom\^{a}nia}
\ead{lpopa@venus.nipne.ro}

\begin{abstract}
For a robust interpretation of upcoming observations
from {\sc Planck}  and  LHC experiments 
it is imperative to understand how the inflationary dynamics of a 
non-minimally coupled Higgs scalar field 
may affect  the degeneracy of the inflationary observables. \\
We constrain the inflationary observables 
and the Higgs boson mass during observable inflation 
by fitting the Hubble function
and subsequently the Higgs inflationary potential directly
to WMAP5+BAO+SNI
We obtain for Higgs mass $m_{Higgs}=143.73^{\,\,\,\,14.97}_{-6.31}$GeV at 95\% CL 
for the central value of top quark mass.
We show that the inflation driven by a non-minimally coupled scalar field to 
Einstein gravity leads to significant changes of the inflationary parameters when 
compared with the similar constraints from the standard inflation.
\end{abstract}

\end{frontmatter}

\section{Introduction}
\label{S1}
The primary goal of particle cosmology is to obtain a concordant description of 
the early evolution of the Universe consistent with both unified field theory 
and astrophysical and cosmological measurements.
Inflation is the most simple and robust theory able to explain the astrophysical and cosmological
observations, providing at the same time self-consistent primordial initial conditions
\cite{Staro80} and mechanisms for
the quantum generation of scalar (curvature) and tensor (gravitational waves)
perturbations \cite{Muk81}.\\
In the simplest class of inflationary models, inflation is driven by a single
scalar field $\phi$ (inflaton) with some potential
$V(\phi)$ minimally coupled to  Einstein gravity. The perturbations are predicted
to be adiabatic, nearly scale-invariant and Gaussian distributed,
resulting in an effectively flat Universe.\\
The WMAP 5-year CMB measurements alone \cite{Dunkeley} or
complemented with other cosmological datasets \cite{Komatsu}
support  the standard inflationary predictions of a nearly flat Universe with adiabatic initial density perturbations. 
In particular, the detected anti-correlations between temperature and E-mode polarization
anisotropy on degree scales \cite{Nolta} provide strong evidence for correlation on
length scales beyond the Hubble radius.

Alternatively one can look on inflationary dynamics based on the 
structure of the Standard Model (SM) of particle physics. 
Recently, Bezrukov and Shaposhnikov \cite{BeSh1} reported 
the  possibility that the SM with an additional non-minimally 
coupled term of the Higgs field and Ricci scalar can give rise to inflation 
without the need for additional degrees of freedom already present in  SM. 
The resultant Higgs inflaton  effective  potential 
in the inflationary domain allows for a range of cosmologically values 
for the Higgs boson mass even after the inclusion of the radiative corrections
\cite{BaSt1,BaSt2,BeSh,dSW,BGS}. 
More precisely, the present cosmological constraints on the Higgs mass 
are based on mapping between the Renormalization Group (RG) flow and 
the spectral index of the curvature perturbations emerging from the analysis 
of WMAP 5-year data complemented with astrophysical distance 
measurements \cite{Komatsu}.\\
However, after the publication of WMAP results, many works devoted to 
the reconstruction of the inflationary dynamics during observable inflation \cite{Julien,Popa} show the compatibility between WMAP results and the predictions of the standard inflation driven by a single scalar field with some potential $V(\phi)$ which is minimally coupled with Einstein gravity.

In order to have a robust interpretation of upcoming observations
from {\sc Planck} \cite{Planck}  and  LHC \cite{LHC} experiments it is imperative to understand 
how the inflationary dynamics of a non-minimally coupled Higgs scalar field 
may affect  the degeneracy of the inflationary observables. \\
In this paper we aim to constrain the inflationary observables 
and the Higgs boson mass during observable inflation 
by fitting the Hubble function $H(\varphi)$, 
and subsequently the Higgs inflationary potential $V(\varphi)$, directly
to WMAP 5-year data \cite{Dunkeley,Komatsu} complemented  with
geometric probes  from the Type Ia supernovae (SN) distance-redshift relation and
the baryon acoustic oscillations (BAO) measurements.

\section{Perturbations driven by non-minimally coupled scalar field}
\label{S2}
We briefly review the computation of the power spectra of 
scalar and tensor density perturbation generated during inflation 
driven by a single scalar field non-minimally coupled to gravity via 
the Ricci scalar {\it \bf R} \cite{FU,HW,KF0,KF,TG}.\\
{\it Jordan frame}. We start with the generalized action in the Jordan frame \cite{FU}:
\begin{equation}
\label{action}
S=\int d^4 x \sqrt{-g} \left[ \frac{1}{2} U(\varphi) {\it \bf R}
            -\frac{1}{2}G(\varphi)(\nabla \varphi)^2 -V(\varphi)\right] \,,
\end{equation}
where $U(\varphi)$ is a general coefficient of the Ricci scalar, 
$G(\varphi)$ is the general coefficient of kinetic energy and $V(\varphi)$ 
is the general potential. 
The generalized $U(\varphi)${\it \bf R} gravity theory in (\ref{action}) 
includes diverse cases of coupling. \\
For generally coupled scalar field $U=(\gamma + \xi m_{Pl}^{-2} \varphi^2)$ and $G(\varphi)=1$ where $m_{Pl} \simeq 2.4 \times 10^{18}$ GeV is the 
reduced Planck mass and $\gamma$ and $\xi$ are constants.
The non-minimally coupled scalar field is the case 
with $\gamma=1$ while the conformal coupled scalar field is the case with 
$\gamma=1$ and $\xi=-1/6$. Throughout the paper we take $sign(\xi)=-1$,  
keeping the same sign convention for $\xi$ as in  Ref. \cite{FU}. 
One should note that other authors \cite{TG,FM} use the opposite 
sign for $\xi$. 
The background equations in a flat Friedmann-Lemaitre-Robertson-Walker 
(FLRW) cosmology are given by:
\begin{eqnarray}
\label{back1}
& & H^2 \equiv \left(\frac{\dot{a}}{a}\right)^2=\frac{1}{6U} 
\left(G \dot{\varphi}^2+ 2U-6H\dot{U}\right)\,,~~~\\
\label{back2}
& & \dot{H}=\frac{1}{2U} \left( -G\dot{\varphi}^2+H\dot{U}
-\ddot{U}\right)\,, \\
\label{back3}
& &\ddot{\varphi}+3H\dot{\varphi}+\frac{1}{2G}
\left(G_{\varphi} \dot{\varphi}^2 - {\it \bf R}U(\varphi)_{,\,\varphi}
 +2 \, V(\varphi)_{,\,\varphi}  \right)=0\,,
\end{eqnarray}
where $a$ is the cosmological scale factor ($a_0=1$ today), 
$H$ is the Hubble expansion rate, overdots denotes the time derivatives and 
$_{,\varphi} \equiv \partial/\partial \varphi $. \\
{\it Einstein frame.}
The conformal transformation to the Einstein frame for the action (\ref{action})  
can be achieved by defining the Einstein metric as \cite{HW}:
\begin{eqnarray}
\label{metric_tr}
\hat{g}_{\mu, \nu}=\Omega g_{\mu \nu}\,,
\end{eqnarray}
where $\Omega=U(\varphi)$ is the conformal factor.
The action in the Einstein frame takes the canonical form:
\begin{eqnarray}
\label{actionE}
S_E=\int d^4 x \sqrt{-{\hat g}} \left[ \frac{1}{2} {\it \bf \hat {R}}
            -\frac{1}{2} (\nabla \hat{\varphi})^2 -\hat{V}(\varphi)\right] \,,
\end{eqnarray}
where the new scalar field $\hat{\varphi}$ and the new potential $\hat{V}$ are defined through:
\begin{eqnarray}
\label{fieldE} 
{\cal S}(\varphi)^{-2} \equiv
\left( \frac{ {\rm d} \hat{\varphi}} {{\rm d}\varphi} \right)^2 =
\left . \frac{3 m^2_{Pl}}{2}\left(\frac{U'(\varphi)}{U(\varphi)}\right)^2+ \frac{G(\varphi)}{U(\varphi)} \right |_{\varphi=\varphi(\hat {\varphi})}
\hspace{0.45cm}\left . 
\hat{V}(\varphi)=\frac{V(\varphi)}{U^2(\varphi)}\right 
|_{\varphi=\varphi(\hat {\varphi})}\,. 
\end{eqnarray}
The Higgs field suppression factor ${\cal S}(\varphi)$ defined in the above equation was first introduced in Ref.\cite{Bond} in the context of CMB anisotropy generation 
in non-minimally inflation model to account for the suppression 
of the Higgs propagator due to non-minimal coupling with gravity.  \\
Decomposing the conformal factor into 
the background part and the perturbed part:
\begin{equation}
\Omega(x,t)=\Omega(t)\left( 1+\frac{\delta \Omega(x,t)}{\Omega(t)} \right)\,,
\end{equation}
we get:
\begin{eqnarray}
\label{transform}
\hat{a}=a \sqrt{\Omega}\,,
\hspace{0.3cm}
{\rm d}\hat{t}=\sqrt{\Omega}{\rm d}t \,,
\hspace{0.3cm}
\hat{H}=\frac{1}{\sqrt{\Omega}} \left( H+\frac{\dot \Omega}{2 \Omega}\right)\,,
\hspace{0.3cm}
\hat{\varphi}=\varphi+ \frac{\delta \Omega} {\Omega} \,.
\end{eqnarray}
In the above relations $\hat{a}$, $\hat{t}$, $\hat{H}$, $\hat{\varphi}$ are respectively the cosmological scale factor, the cosmic time, the Hubble expansion rate and the scalar field in the Einstein frame.

\subsection{Cosmological perturbations in the Jordan frame}

{\it Scalar perturbations} (S). The equation of motion for the Lagrangian (\ref{action})
is given by \cite{HW}:
\begin{eqnarray}
\label{eq_motion}
 \frac{1}{a^3Q_{\rm S}}  \left(a^3 Q_{\rm S} \dot{\cal R}\right)^{\bullet}
+\frac{k^2}{a^2}{\cal R}=0\,,~~~{\rm with}~~~
Q_{\rm S}=\frac{\dot{\varphi}^2+3\dot{U}^2/2U}
{(H+\dot{U}/2U)^2}\,.
\end{eqnarray}
Here $k$ is the comoving and ${\cal  R}$ is the intrinsic curvature perturbation of the 
co-moving hypersurfaces during inflation: 
\begin{equation}
\label{curvature}
{\cal R} = \psi -\frac{H}{\dot \varphi} \, \delta \varphi \,,
\end{equation}
where $\delta \, \varphi$ is the perturbation of the scalar field $\varphi$.
Neglecting the contribution of the decaying modes, 
the scale dependence of the amplitudes of the scalar 
perturbations can be obtained  by integrating the mode  
equation  \cite{MU}:
\begin{eqnarray}
\label{Muk}
u_k''+ \left(k^2- \frac{z''}{z} \right)u_k  =  0 \,.
\end{eqnarray}
In the above equation the prime denote the  derivative with respect to the conformal time $\eta(a)=\int a(t)^{-1}dt$. For scalar perturbations the term $z''/z $ can be written as \cite{TG}:
\begin{eqnarray}
\label{zS}
\frac{z''_{\rm S}}{z_{\rm S}}= 
(aH)^2 \left[ (1+\delta_{\rm S}) (2+\delta_{\rm S} + \epsilon) + 
\frac{\delta'_{\rm S}}{aH} \right]\,,
\end{eqnarray} 
where
\begin{eqnarray}
\label{JE}
z_{\rm S}=a \sqrt{ Q_{\rm S} }\,, 
\hspace{0.5cm} 
\epsilon=\frac{ {\dot H} }{H^2}\,, 
\hspace{0.5cm} 
\delta_{\rm S}= \frac{ {\dot Q}_{\rm S}} {2 H Q_{\rm S} } \,.
\end{eqnarray}
{\it Tensor perturbations} (T). Tensor perturbations satisfy the same form of 
equation as Eq.(\ref{eq_motion}) with replacement $Q_S \rightarrow Q_T=U$ \cite{FU}.
The scale dependence of the amplitude of tensor (T)
perturbations can also be obtained  by integrating  Eq.(\ref{Muk}) 
taking the term $z''/z $ of the form \cite{TG}:
\begin{eqnarray}
\label{zT}
\frac{z''_{\rm T}}{z_{\rm T}}= 
(aH)^2 \left[ (1+\delta_{\rm T}) (2+\delta_{\rm T} + \epsilon) + 
\frac{\delta'_{\rm T}}{aH} \right]\,,
\end{eqnarray}
where
\begin{eqnarray}
z_{\rm T}=a \sqrt{ Q_{\rm T} }\,, 
\hspace{0.5cm} 
\delta_{\rm T}= \frac{ {\dot Q}_{\rm T}} {2 H Q_{\rm T} } \,.
\end{eqnarray}

\subsection{The equivalence of the perturbations spectra}

Introducing the following quantities: 
\begin{eqnarray}
\hat{\epsilon} & = & \frac{d\hat{H}}{d\hat{t}} \,,
\hspace{0.6cm}
\hat{\delta_{\rm S}}=\frac{d \hat{Q}_{\rm S}/d\hat{t}}{2 \hat{H} \hat{Q_{\rm S}}}\,,
\hspace{0.6cm}
\hat{\delta_{\rm T}}=\frac{d \hat{Q}_{\rm T}/d\hat{t}}{2 \hat{H} \hat{Q_{\rm T}}}\,, \\
\hat{Q}_{\rm S} & = & \left(\frac{d \hat{\varphi}}{\hat {H}}\right)^2=\frac{Q_{\rm S}}{U} \,,
\hspace{0.9cm}
\hat {Q}_{\rm T}=\frac{Q_{\rm T}}{U} \,, \nonumber
\end{eqnarray}
and making use of the relations given in Eq.(\ref{JE}), it  can shown that both scalar and tensor perturbations in 
the Jordan frame exactly coincide with those in the Einstein frame 
\cite{CH,Staro1}
implying that power spectra of the cosmological perturbations are invariant 
under the conformal transformation:
\begin{eqnarray}
\hat{\cal P}_{\rm S}(k)={\cal P}_{\rm S}(k)\, , \hspace{0.8cm} 
\hspace{0.4cm}
\hat{\cal P}_{\rm T}(k)={\cal P}_{\rm T}(k)\,.
\end{eqnarray}
 
The numerical evaluation of the perturbations power spectra involves solving 
equations (\ref{Muk}), (\ref{zS}) and (\ref{zT}) for each value of the wavenumber $k$, 
the evolution of  $|u_k| /z_{S,T}$ to a constant value defining the observable power spectra ${\cal P}_{{\rm S,T}}$. Then
the scalar and tensor spectral indexes $n_{\rm S,T}$ and the running of scalar spectral index
$\alpha_S$ at the Hubble radius crossing $k_*$ are defined as:
\begin{eqnarray}
\label{inflpar_e}
n_{\rm S}-1   \equiv  \left.  
\frac{d\,{\rm ln}{\cal P}_{\rm S}(k)}{ d\,{\rm ln}k} \right |_{k_*=aH}
\hspace{0.4cm}
n_{\rm T}   \equiv   \left. \frac{d\,{\rm ln}{\cal P}_{\rm T}(k)}{d\,{\rm ln}k} \right |_{k_*=aH}
\hspace{0.4cm}
\alpha_{\rm S} \equiv  \left. \frac{d \,n_{\rm S}(k)}{d\,{\rm ln}k} \right|_{k_*=aH}
\end{eqnarray}

\section{Higgs boson as inflaton}

{\it Higgs field couplings in the inflationary domain}. 
Higgs boson as inflaton add non-minimal coupling to gravity \cite{FM,BaSt1,BaSt2,BeSh,dSW}. Taking Higgs doublet in the form $(1/\sqrt{2})(0, {\it v} +\varphi)$, the part of the general action given in Eq.(\ref{action}) relevant for inflation is only the scalar part.
The quantum corrections due to the effects of particles of 
the Standard Model interacting with the Higgs boson
through quantum loops  modify the action coefficients $U$, $G$ and $V$ 
that in the Jordan frame  take the form \cite{dSW,BBKKS}:
\begin{eqnarray}
\label{UH}
U(t) & = & 1+ \xi(t)G^2(t) \frac{\varphi(t)^2}{m_{Pl}^{-2}}  = 
 1+  \xi(t)G(t)\frac{m^2_T}{m^2_{Pl}} e^{2t}\,,\\
\label{GH}
G(t) & = & {\rm exp} \left(\int^{t}_{0} dt'\gamma(t')\right) \,, \\
\label{VH} 
V(t) & = & \frac{\lambda(t)}{4}G^4(t) ((\varphi^2(t)^2-{\it v}^2)^2 
\simeq  \frac{\lambda(t)}{4}G^4(t)m^4_T \,e^{4t}\,,
\end{eqnarray}
where ${\it v} \simeq 246.2$ GeV is the vacuum expectation value for the Higgs, 
$\lambda$ is the quadratic coupling constant of the Higss 
boson with the mass $m_H=\sqrt{2 \lambda}{\it v}$ and $\gamma(t)$ is the Higgs field anomalous dimension.  
In the above equations the scaling variable
\begin{eqnarray}
\label{scaling}
t=t(\varphi)= {\rm ln}(\varphi/m_T)\,,\hspace{0.45cm}\varphi(t) =m_T e^t
\end{eqnarray}
normalizes the Higgs field and all the running couplings 
to the top quark mass scale $m_T=171.3 \pm 1.1 \pm 1.2$ GeV \cite{PDG}.
 As the energy scale of inflation is many order of magnitude above the electroweak scale ($\varphi(t) >> {\it v}$), in the right hand side of Eq.(\ref{VH}) the potential was approximated by a quadratic potential $V(\varphi)=\lambda \varphi^4/4$ \,.\\ 
The Higgs field  suppresion factor ${\cal S}(t)$,  the Higgs potential  $\hat{V}(t)$ 
and the first two potential slow-roll (PSR) parameters $\epsilon_{V}$ and $\eta_V$ in 
the Einstein frame  are given by:
\begin{eqnarray}
\label{higgsV}
{\cal S}^2(t)=\frac{2}{m^2_{Pl}}\frac{U^2(t)}{G(t)U(t)+3U^2_{,\,\varphi(t)}} \,,
\hspace{0.45cm}
\hat{V}(t)= \frac{ V(t)} {U^2(t)}  \,, 
\end{eqnarray}
\begin{eqnarray}
\label{PSR} 
\epsilon_V=\frac{1}{2}m^2_{Pl}\left( \frac{V_{,\,\varphi(t)}}{V}
\right)^2\,{\cal S}^{-2}\,,
\hspace{0.45cm}
\eta_V=2\epsilon+m_{Pl}\frac{\epsilon_{V_{,\,\varphi(t)}}}{\sqrt{2 \epsilon_V}}\, {\cal S}^{-1}\,.
\end{eqnarray}
The amplitude of the scalar (curvature) perturbations at the Hubble radius crossing $k_*$ 
is related to the Higgs potential in the Einstein frame through:
\begin{eqnarray}
\label{As}
A^2_S = \left. \frac{\hat V}{24 \pi^2 m^4_{Pl} \epsilon_V} \right |_{k_*} \,.
\end{eqnarray}
We compute the various $t$-dependent running constants, the Higgs field propagator suppretion factor and  Higgs field anomalous dimension  
by integrating the two-loop renormalization group 
$\beta$-functions for $SU(2)\times S(1)$ gauge couplings ${g',g}$, the $SU(3)$ strong coupling $g_s$, the top Yukawa coupling $h_t$ and the Higgs 
quadratic coupling $\lambda$, as collected in the Appendix of \cite{Riotto}.
For the non-minimal coupling constant $\xi$ we use the $\beta$-function 
from the Appendix of \cite{Clark}.
At $t=0$, which corresponds to the top quark mass scale,  
the top Yukawa coupling $h_t(0)$ and Higgs quadratic coupling $\lambda(0)$ 
are fixed by their pole masses \cite{Riotto}. The gauge coupling constants 
at $m_T$ scale are taken from \cite{BaSt2}: $g^2(0)=0.4202$, $g'^2 (0)=0.1291$ and $g^2_s (0)=0.1.3460$. 
The value of $\xi(0)$ is determined so that at the initial point of the slow roll inflation, $t_i$, the non-minimal coupling constant $\xi(t_i)$ is such that the calculated value of the amplitude of density perturbations given in Eq.(\ref{As}) agrees with the measured result 
obtained by WMAP team \cite{Komatsu}.

{\it Higgs field equations of motion.} 
Hereafter we will work in the Einstein frame dropping the `hat' when  express 
different quantities in this frame. The inflationary dynamics and the cosmological perturbations power spectra are entirely determined by the dynamics of the Higgs field $\varphi(t)$. \\
By using the transformation relations from Jordan frame to Einstein frame  given in  Eq.(\ref{transform}), to the leading order in slow-roll approximation ($\ddot \varphi<< 3 H \varphi$, neglecting also the higher powers in $\dot \varphi$ ) 
the general background Eqs. (\ref{back1}) - (\ref{back3}) reduce to \cite{KF,TG}:. 
\begin{eqnarray}
\label{Hf1}   
H^2 &  \equiv & \left( \frac{\dot a}{a}\right)^2 \simeq
\frac{1}{3m_{Pl}^{2}} \left[\frac{\dot \varphi^2}{U}\,{\cal S}^{-2}  + V \right] \,, \\ 
\label{Hf2}
3 H {\dot \varphi} & \simeq & - U V_{,\,\varphi(t)} \,\,{\cal S}^{-2} \,.
\end{eqnarray}
It follows that the amount of the expansion from any epoch to the end of the inflation 
($\epsilon_V=1$) that is controlled by the number of {\it e}-folds ${\cal N}$ is:
\begin{equation}
\label{efold}
{\cal N}=\int^{\varphi_{end}(t)}_{\varphi(t)} d\varphi'(t) 
\frac{3\,H^2(\varphi'(t))}{F(\varphi'(t))} \,.
\end{equation}

{\it The CMB angular power spectra.} 
We obtain the CMB temperature anisotropy and polarization power spectra 
by  integrating the mode equation (\ref{Muk}) and the Higgs field equations 
(\ref{Hf1}) and  (\ref{Hf2}) with respect to the conformal time. 
We consider wavenumbers in the range [$5 \times 10^{-6}-5$]~Mpc$^{-1}$
needed to numerically derive the CMB angular power spectra
and the Hubble crossing  scale $k_*=0.002$Mpc$^{-1}$.
At this scale,  the value of the 
Higgs scalar field  $\varphi_*$  corresponds to 
${\cal N}=k_*/aH\simeq$ 59 ${\it e}$-folds till the end of the inflation 
and the value of the scaling parameter is $t_* \simeq$37 \cite{BaSt1,BaSt2}, 
as indicated by the central value of the amplitude of scalar perturbations 
measured by WMAP at $k_*$ \cite{Komatsu}.
For each wavenumber $k$ in the above range our code integrates 
Eqs.(\ref{Muk}), (\ref{Hf1}) and  (\ref{Hf2})
and the $\beta$ functions of the  $t$-dependent running constant couplings 
in the observational inflationary window imposing 
that $k$ grows monotonically to the wavenumber $k_*$, 
eliminating at the same time the models violating the condition for 
inflation $0 \le \epsilon_H \equiv -{\dot H}/H^2 \le 1$. \\
We reconstruct the Hubble expansion rate $H(\phi-\phi_*)$
from the data by using the Taylor expansion up to the cubic term, 
equivalent to keeping the first three slow-roll parameters. \\
For comparison, we compute the CMB angular power spectra 
in the standard inflation model with  minimally coupled scalar field 
with a general potential $V(\phi)$. The details of this computation can be found in \cite{Popa}.
\newpage
\section{Results}

\begin{figure}
\includegraphics[height=7cm,width=14cm]{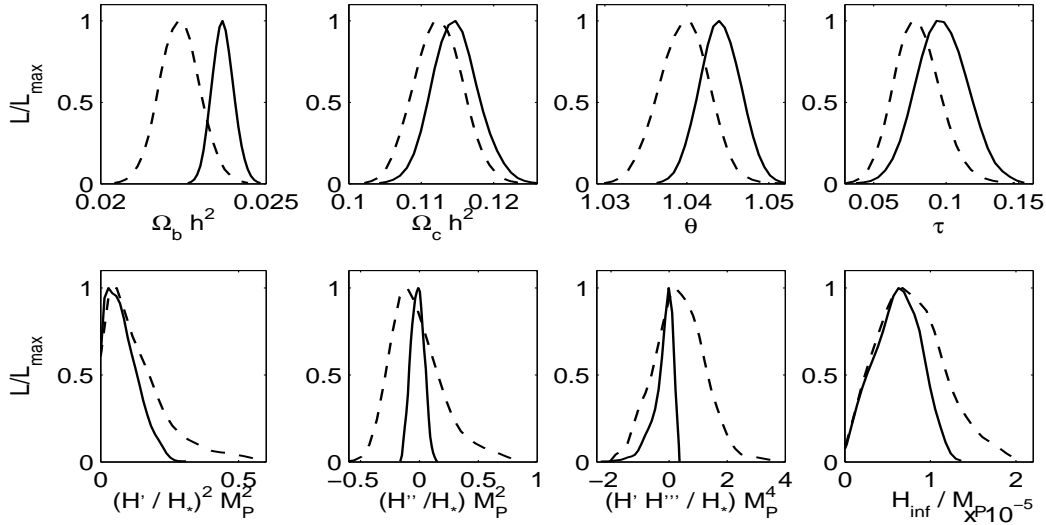}
\caption{1D marginalized likelihood probability distributions 
for seven independent parameters obtained from the fit to WMAP5+SN+BAO dataset of the inflation model 
with non-minimally coupled Higgs scalar field (continuous lines) and of the standard inflation model with minimally coupled scalar field (dashed lines).
The last plot shows the value of Hubble parameter. 
All parameters are computed at the Hubble radius crossing $k_*$=0.002Mpc$^{-1}$.}
\end{figure}
\begin{figure}
\includegraphics*[height=14cm,width=14cm]{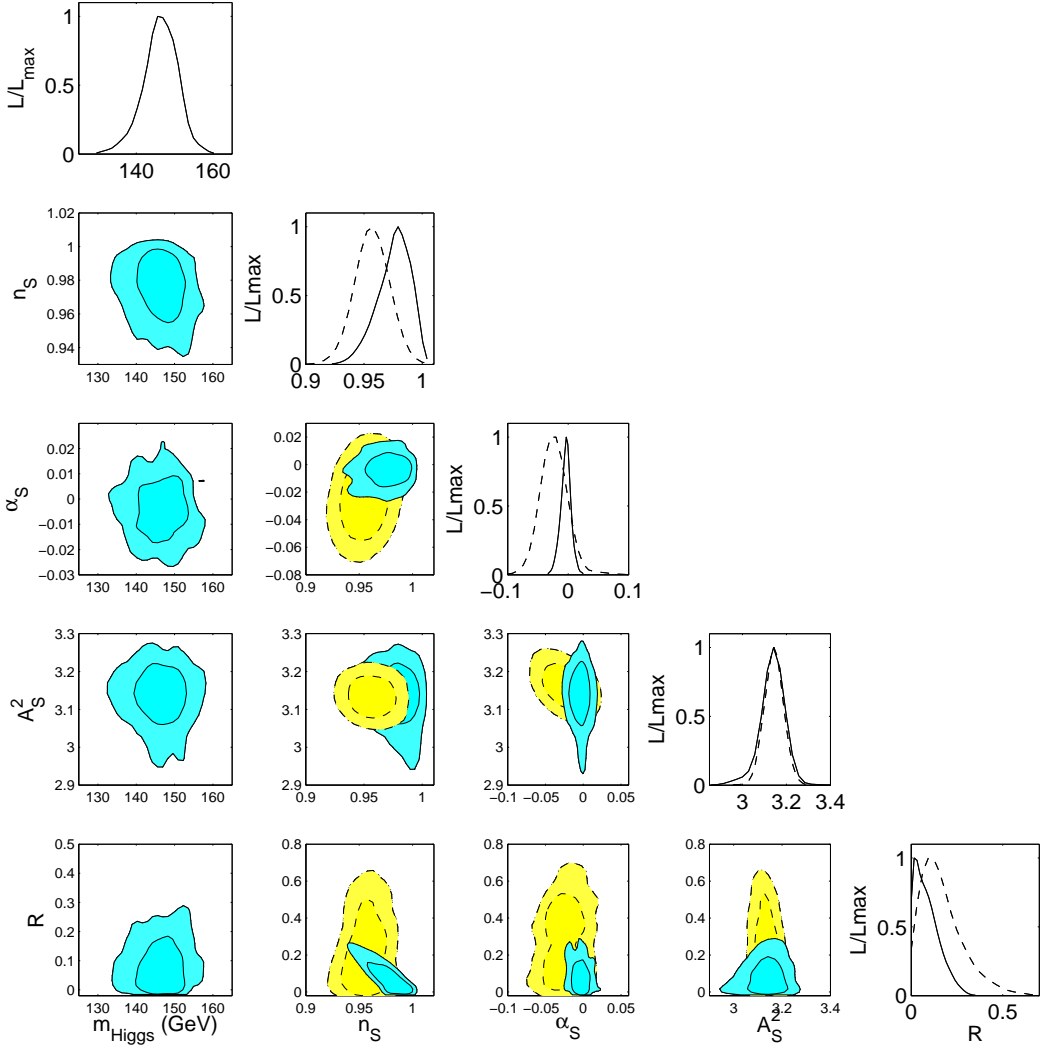}
\begin{center}
\caption{The figure presents the results of the fit to  WMAP5+SN+BAO dataset of
the inflation model with non-minimally coupled Higgs field in blue (continuous lines) 
and the results of the fit to the same dataset of the standard inflation model with 
minimally coupled scalar field in yellow (dashed lines).  
The top plot in each column shows the
probability distribution  of different scalar inflationary observables
while the other plots show their joint 68\% and 95\% confidence intervals.
All parameters are computed at the Hubble radius crossing $k_*$=0.002Mpc$^{-1}$.}
\end{center}
\end{figure}
\begin{figure}

\vspace{-1cm}
\includegraphics*[height=7cm,width=14cm]{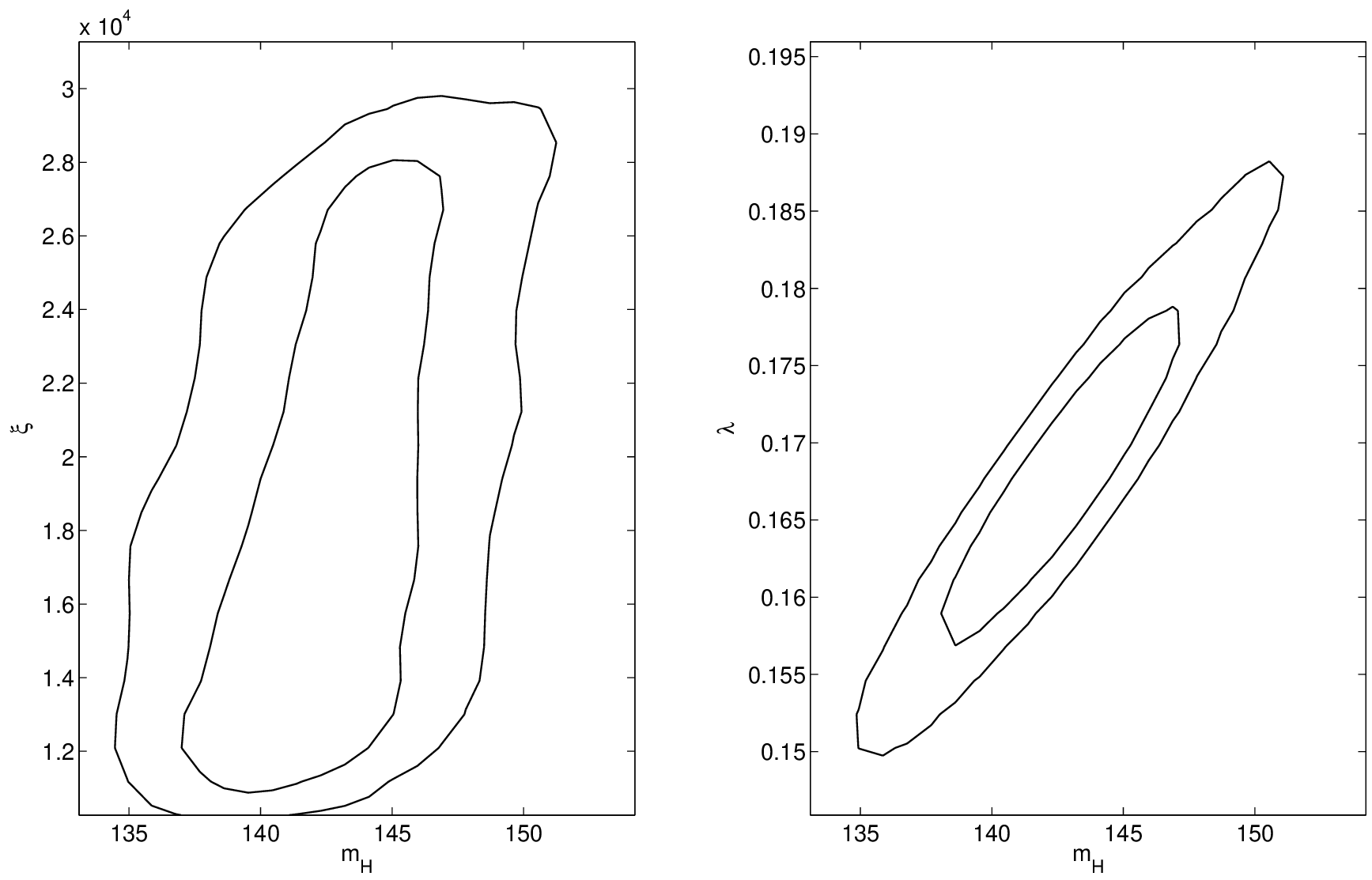}
\caption{2D marginalized likelihood probability distributions  
in the planes $m_{Higgs}$ - $\xi$ (left panel) and $m_{Higgs}$-$\lambda$ (right panel)
obtained from the fit to WMAP5+SN+BAO dataset of the inflation model 
with non-minimally coupled Higgs scalar field.}
\end{figure}
We use the Markov Chain Monte Carlo (MCMC) technique to reconstruct the Higgs field 
potential and to derive constraints on the inflationary observables and the Higgs mass 
from the following datasets.\\
The WMAP 5-year data \cite{Komatsu} complemented  with
geometric probes  from the Type Ia supernovae (SN) distance-redshift relation and
the baryon acoustic oscillations (BAO). The SN distance-redshift relation has been studied in detail in the recent
unified analysis of the published heterogeneous SN data sets -
the Union Compilation08  \cite{Kowalski}. 
The BAO in the distribution
of galaxies  are extracted from the Sloan Digital Sky Surveys (SDSS) and Two Degree Field Galaxy Redshidt Survey (2DFGRS) \cite{Percival}.
The CMB, SN and BAO data (WMAP5+SN+BAO) are combined by multiplying the likelihoods.\\
We decided to use these measurements especially because we are testing models deviating from  the  standard Friedmann expansion.
These datasets properly enables us to account for any shift of the CMB
angular diameter distance  and of the expansion rate of the Universe.

The likelihood probabilities are evaluated  by using
the public packages {\sc CosmoMC} and {\sc CAMB}
\cite{Lewis,camb} modified to include the 
formalism for inflation driven by  non-minimally coupled Higgs scalar field, 
described in the previous sections. 
Our fiducial model is the $\Lambda$CDM standard cosmological model 
described by the following set of parameters receiving uniform priors:
$$ \Omega_bh^2 \,,\,\,\Omega_ch^2\,,\,\, \theta_s\,,\,\, \tau\,,\,\,
A^2_S\,,\,\, m_{Higgs} \,,\,\, \frac{H'^2}{H^2}\,,\,\, \frac{H''}{H}\,,\,\, \frac{H'''H'}{H^2}\,$$
where: $\Omega_{b}h^2$ is the 
physical baryon  density, $\Omega_{cdm}h^2$ 
is the physical dark matter density,  $\theta_s$
is the ratio of the sound horizon distance 
to the angular diameter distance,  $\tau$ is 
the reionization optical depth, $m_{Higgs}$ is the Higgs boson mass and $H'$, $H''$ and $H'''$ are the derivatives of the Hubble expansion rate with respect to the Higgs scalar field. \\
For comparison, we use the MCMC technique to reconstruct the standard inflation field 
potential and to derive constraints on the inflationary observables 
from the fit to WMAP5+SN+BAO dataset of the standard inflation model 
with minimally coupled scalar field and a general potential $V(\phi)$ \cite{Popa}. 
For this case we use the same set of input parameters with uniform priors 
as in the case of non-minimally coupled Higgs scalar field inflation, 
except for Higgs mass. \\
For each inflation model we run 120 Monte Carlo chains, 
imposing for each case the Gelman \& Rubin convergence criterion \cite{Gelman}.

Fig.~1 presents the 1D marginalized likelihood probability distributions 
of the fiducial model parameters obtained from the fit to WMAP5+SN+BAO dataset of the inflation model with non-minimally coupled Higgs scalar field compared
with the similar distributions obtained from the fit to same dataset 
of the standard inflation model with minimally coupled scalar field.
One can see the differences between the likelihood probabilities caused 
by the difference in the dynamics of the scalar fields during inflation.
One should note the smaller value of the Hubble expation rate during inflation for 
the inflation model involving the Higgs field as inflaton.\\
In Fig.~2 we present the constraints on the inflationary parameters, namely 
the scalar spectral index $n_S$, the running of the scalar spectral index $\alpha_S$, 
the amplitude of the scalar (curvature) density perturbations $A^2_S$ and 
the ratio of tensor to scalar amplitudes $R$, as obtained  from  the fit to  WMAP5+SN+BAO dataset of the inflation model with non-minimally coupled Higgs field compared with   
the constrains obtained from the fit to the same dataset of the standard inflation model with minimally coupled scalar field. We also show the constraints obtained on the Higgs mass 
from our analysis. The mean values and the 95\% CL intervals of these parameters are presented in Table~1.\\
We obtain a Higgs mass value of $m_{Higgs}=143.73^{\,\,\,\,14.97}_{-6.31}$GeV at 95\% CL for the central value of top quark mass. \\
Our result is compatible with the previous constrains of the Higgs mass from cosmological data \cite{BaSt1,BaSt2,BeSh,dSW,BGS} however the upper and lower limits 
of the Higgs mass obtained from this analysis are tightly constrained.\\
Table~1 clearly show the differences between the estimates of the inflationary 
parameters obtained from the fits of the inflation models.\\ 
Fig~3 shows the 2D marginalized likelihood probability distributions  
in the planes $m_{Higgs}$ - $\xi$ and $m_{Higgs}$-$\lambda$. 
One should note the strong correlation between Higgs mass and 
the Higgs quadratic coupling $\lambda$ and the larger degeneracy between the Higgs mass 
and non-minimal coupling constant $\xi$, indicating 
the existence of a degeneracy between $\xi$ and $\lambda$ that affect the Higgs mass 
value inferred from cosmological and astrophysical measurements. 

\section{Conclusions}

In order to have a robust interpretation of upcoming observations
from {\sc Planck} \cite{Planck}  and  LHC \cite{LHC} experiments 
it is imperative to understand how the inflationary dynamics of a 
non-minimally coupled Higgs scalar field 
may affect  the degeneracy of the inflationary observables. \\
We constrain the inflationary observables 
and the Higgs boson mass during observable inflation 
by fitting the Hubble function $H(\varphi)$, 
and subsequently the Higgs inflationary potential $V(\varphi)$, directly
to WMAP 5-year data complemented  with
geometric probes  from the Type Ia supernovae (SN) distance-redshift relation and
the baryon acoustic oscillations (BAO) measurements.
We obtain a Higgs mass value of $m_{Higgs}=143.73^{\,\,\,\,14.97}_{-6.31}$GeV at 95\% CL for the central value of top quark mass $m_{Top}$=171.3 GeV. 
Our result is compatible with the previous constrains of the Higgs mass from cosmological data \cite{BaSt1,BaSt2,BeSh,dSW,BGS}, 
however the upper and lower bounds of the Higgs mass obtained 
from our analysis are tightly constrained.

The strong correlation between Higgs mass and 
the Higgs quadratic coupling $\lambda$ and the larger degeneracy between the Higgs mass 
and non-minimal coupling constant $\xi$, indicate
the existence of a degeneracy between $\xi$ and $\lambda$ 
that affect the Higgs mass value inferred from cosmological and astrophysical measurements. 

We also show that the inflation driven by a non-minimally coupled scalar field to 
Einstein gravity leads to significant changes of the inflationary parameters when 
compared with the similar constraints from the standard inflation with minimally coupled 
scalar field.

\vspace{1cm}
{\bf Acknowledgments}\\ \\
The author acknowledge D. Ghilencea for usefull discussions.\\
This work  was partially supported by CNCSIS Contract 539/2009.

\begin{table}
\caption{The mean values and 95\% CL lower and upper intervals
of the derived parameters obtained from the fit of the inflation model 
with non-minimally coupled Higgs scalar field 
and the standard inflation model with minimally coupled scalar field
to WMAP5+SN+BAO dataset.
All parameters are computed at the Hubble radius crossing $k_*$=0.002 Mpc$^{-1}$.}
\begin{center}
\begin{tabular}{lcc}
\hline \hline \\
Model & Standard Inflation   &  Higgs Inflation  \\
Parameter &                  &                   \\
\hline \\
$\Omega_bh^2$&$0.022_{0.021}^{0.023}$&$0.023_{0.022}^{0.024}$   \\ 
$\Omega_ch^2$& $0.111_{0.106}^{0.117}$  &$  0.114_{0.107}^{0.121}  $\\         
$\tau$       & $0.082_{0.055}^{0.109}$  &$   0.094_{0.034}^{0.135} $\\ 
$\theta_s$   & $1.039_{1.034}^{1.045}$  &$ 1.044_{1.037}^{1.048} $   \\ 
${\rm ln}[ 10^{10}A^2_S ]$ & $3.143_{3.083}^{3.201}$  & $3.317_{2.999}^{3.252}$   \\  
$n_S$ &  $0.956_{0.932}^{0.979} $ & $ 0.977_{0.944}^{0.997}         $            \\  
$\alpha_S$   &   $-0.012_{-0.038}^{\,\,\,\,0.021}$ &$0.003_{-0.0021}^{\,\,\,\,0.0133}$ \\ 
$R$ &     $< 0.556$         & $< 0.234$                 \\
\hline
$m_{Higgs}$(GeV)&  &$142.737_{136.431}^{158.707}$\\ 
$\lambda$   &  &$0.168_{0.153}^{0.185}$\\ 
$\xi \times 10^{-5}$& &  $1.871_{0.118}^{0.280}$\\ 
\hline \hline
\end{tabular}
\end{center}
\end{table}

\end{document}